\documentclass[aps,prl,reprint,groupedaddress,showpacs]{revtex4-1}

\usepackage{graphicx}
\usepackage{amsmath}
\usepackage{textcomp}
\usepackage{color}

\begin{document}

\title{Edge states protected by chiral symmetry in disordered photonic graphene}

\author{J. M. Zeuner$^1$, M. C. Rechtsman$^2$, S. Nolte$^1$, and A. Szameit$^1$}
\email{alexander.szameit@uni-jena.de}

\affiliation{$^1$Institute of Applied Physics, Friedrich-Schiller-University Jena, Max-Wien-Platz 1, 07743 Jena, Germany}

\affiliation{$^2$Solid State Institute and Physics Department, Technion, 32000 Haifa, Israel}

\date{\today}

\begin{abstract}
We experimentally investigate the impact of uncorrelated composite and structural disorder in photonic graphene. We find that in case of structural
disorder not only chiral symmetry, but also the vanishing of the density of states at zero energy is preserved. This is in contrast to composite
disorder, where chiral symmetry as well as the vanishing of the density of states are destroyed. Our observations are experimentally proven by
exciting edge states at the bearded edge in disordered photonic graphene.
\end{abstract}

\maketitle

Graphene, a honeycomb patterned monolayer of carbon atoms, has been celebrated as a new wonder material since its first experimental
realization in 2004 \cite{Novolesov:Science2004}. Especially due to its unique properties such as, e.g., very high thermal conductivity, high
charge-carrier mobility and very high mechanical stability, this material is a candidate for many different future electronic and photonic
devices. Due to its special properties, graphene offers a platform for addressing a number of questions of a fundamental physical nature. Indeed,
novel phenomena such as the (fractional) quantum Hall effect \cite{Zhang:Nature2005,Novoselov:Nature2005}, strong suppression of quantum interference
\cite{Morozov:PRL2006}, and Klein tunnelling \cite{Katsnelson:NaturePhysics2006} have been shown to exist in this material. The graphene lattice has
a two-member basis, composed of shifted hexagonal lattices, and can be therefore described in a tight-binding model containing two bands. The bands
meet at conical intersections (called Dirac points) at zero energy, where electrons have a dispersion relation akin to that of the Dirac equation.
Importantly, the density-of-states (DOS) near those points goes to zero. If only nearest-neighbor atoms interact, atoms of one sublattice interact
merely with the other sublattice and not their own. Such systems possess chiral symmetry \cite{Wallace, Haldane, Hatsugai2006, Hatsugai2007,
GrapheneReview}, from which two important properties follow: (1) any state with nonzero energy $E$ must have a chiral partner with energy
$-E$ and (2) any state with
zero energy cannot be altered by perturbations respecting the chiral symmetry.

Due to the fundamental difference of Dirac-like dispersion and Schr\"odinger-like dispersion, a full understanding of the impact of disorder on
graphene has been recently extensively investigated, but a consensus has not been fully reached \cite{Altland, Merlin,
DasSarma:RevModPhys2011,Mucciolo:JPhys2010}. Disorder that is introduced into the system may be categorized in two different ways: composite disorder
(also called on-diagonal disorder), which corresponds to impurities in the lattice, and structural disorder (also called off-diagonal disorder),
which corresponds to random displacements of the atoms in the lattice. Composite disorder destroys the chirality and therefore the Dirac spectrum with the associated vanishing DOS at zero energy \cite{Leconte:PRB2011,Merlin, Wimmer:PhysRevB2010}. Structural disorder preserves the chiral symmetry of the honeycomb lattice \cite{Merlin}. However, how this affects transport properties and the DOS in the Dirac region of the spectrum is still an open question.

\begin{figure}[htb]
\includegraphics[width=8cm]{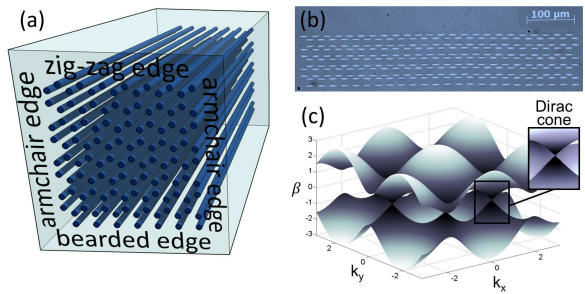}
\caption{(a) Scheme of the waveguides arranged in a honeycomb structure, the so-called photonic graphene with the labelling of the different edge
types. (b) Microscopic image of the structure. (c) Dispersion relation of photonic graphene which equals that of electronic
graphene. \label{fig:setup}}
\end{figure}

In this work, we experimentally probe the impact of composite and structural disorder in a graphene-like lattice, and demonstrate the physical
effects for both cases, chiral symmetry breaking and preserving. We present strong experimental evidence that under structural disorder - which preserves the chiral symmetry - the vanishing DOS at the Dirac point remains.
To this end we use optical honeycomb waveguide lattices - \textit{photonic graphene} - which constitute an exceptional system for probing the unique graphene geometry
\cite{Peleg:PRL2007,Rechtsman:NatPhoton2012}. We use an optical system since probing the effects of disorder in atomic graphene is an essentially
impossible task due to the difficulty of engineering graphene samples with controllable amounts of disorder, types thereof, and correlations therein.
In our system, we employ zero-energy edge states at the bearded edge of the lattice \cite{Kohmoto:PhysRevB2007} which turn out to be well suited for
experimentally analyzing the spectrum at the zero-energy region. We find that these edge states are destroyed under composite disorder, but survive
structural disorder. This suggests that the vanishing of the DOS at zero energy is preserved in the latter case. Note that such edge states were experimentally demonstrated only
recently in photonic graphene \cite{Rechtsman:arxiv2012} as in electronic graphene an expanded bearded edge was not realized so far due to stability
problems \cite{Liu:PRL2009}. A very interesting feature of such edge states was
demonstrated lately: In  \cite{strainedHC, Bellec:PRL2013} it was shown that edge states
experience a topological transformation when a linear strain is applied to
the lattice. In Fig. \ref{fig:setup}(a), a sketch of a honeycomb photonic lattice with all basic edge types is depicted. A micrograph
of the front facet of one fabricated sample is shown in Fig. \ref{fig:setup}(b) with the bearded edge at the bottom.

\begin{figure*}[htb]
\includegraphics[width=.72\textwidth]{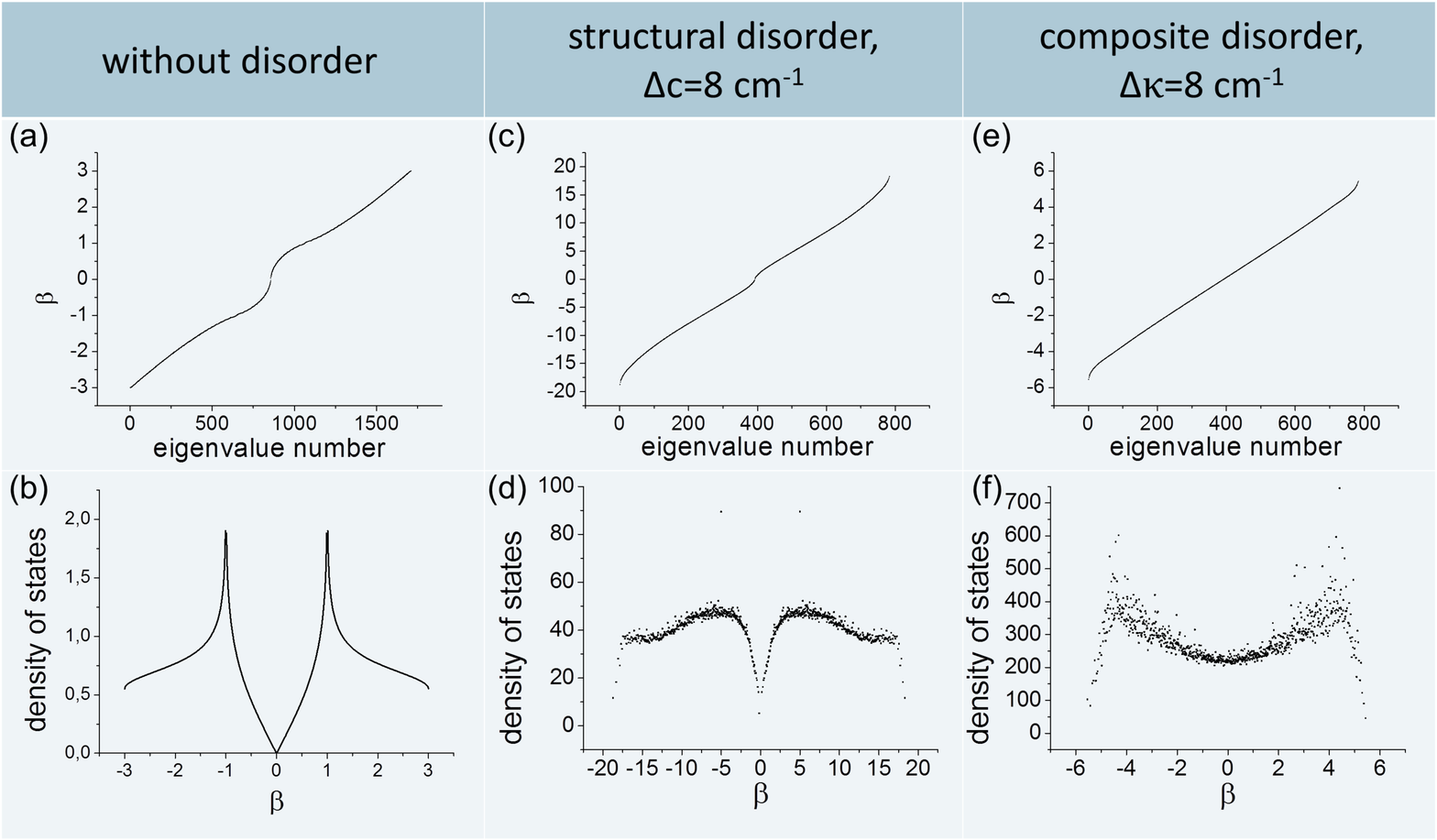}
\caption{Eigenvalues and DOS plotted for three different cases: no disorder, structural and composite
disorder.\label{fig:eigval} }
\end{figure*}

In order to model an optical honeycomb waveguide lattice, we employ the tight-binding-equations
\begin{equation} \label{TBM}
i\partial_z\psi_n(z) = -\sum_{<m>}c_{n,m}\psi_m - \kappa_n\psi_n(z) \equiv H_{TB}\psi_n(z)\;,
\end{equation}
where the summation is taken only over nearest-neighbor sites. Here, $\psi_n$ is the amplitude of the guided mode in the n$th$ waveguide, $z$ is the
propagation direction, $c_{n,m}$ is the coupling constant between the n$th$ and m$th$ waveguide, and $\kappa$ denotes the propagation constant.
Evidently, this can be written using a tight-binding Hamiltonian matrix $H_{TB}$, which is identical to that of electronic graphene. The only
difference of both systems is the evolution coordinate: whereas in electronic graphene the wave function evolves in time, in the optical system the
light propagation occurs along $z$. Hence, the temporal dynamics of electrons in graphene is emulated by light diffraction in the optical waveguide
system. In order to achieve time-independent equations for the eigenvalues $\beta$, one substitutes the plane wave solutions $\psi(z)=\psi
e^{i\beta z}$ into Eq.~\eqref{TBM}, which results in an eigenvalue equation for the eigenmodes of the system:
\begin{equation} \label{EVE}
-\beta \psi = H_{TB}\psi\;.
\end{equation}
For periodic boundary conditions, this equation allows the determination of the band structure of a honeycomb lattice, where $\beta$ is a function of
the transverse Bloch wave numbers $\mathbf{k}=(k_x,k_y)$, as shown in Fig.~\ref{fig:setup}(c). Indeed, there is no bandgap, as both bands intersect
at six singular points. In the vicinity of these Dirac points, the Hamiltonian can be approximated by the relativistic Dirac Hamiltonian
$H_{\mathrm{D}} \sim \mathbf{k} \cdot \mathbf{\sigma}$ with $\mathbf{\sigma}$ being a vector of the Pauli matrices, which is an analogue to the Dirac
equation for massless relativistic particles \cite{Geim:NatMat2007}. This is the central characteristic of the graphene structure and one of the main reasons for many
exotic properties of this material. A further key feature of graphene is the observation that at the Dirac points the DOS vanishes. In
Fig.~\ref{fig:eigval}(a) we show all eigenvalues $\beta$ in ascending order of a system, where periodic boundary conditions are applied in all
directions. The analytic expression of the DOS - which is inversely proportional to the slope of the eigenvalue function shown in
Fig.~\ref{fig:eigval}(a) - is plotted in Fig.~\ref{fig:eigval}(b) and exhibits a clear drop at $\beta=0$. The question arises how stable the
vanishing of the DOS is under the influence of disorder.

We start our analysis using simulations and examine a photonic graphene structure under the influence of both different kinds of disorder: composite
disorder (which manifests in random depth of the lattice sites) and structural disorder (which results by a random shift of the lattice sites around
their average position). Note that in our optical system, we can realize both disorders independently of each other. Moreover, the structural
disorder is fully uncorrelated, in contrast to electronic graphene, where a random shift of the atoms always influences the neighboring sites. We
implement the structural disorder in Eq.~\eqref{TBM} by $c=c_0+\frac{\Delta c}{2}\cdot\xi$, where $c_0$ is the mean coupling constant, $\Delta c$
defines the strength of the disorder in units of the coupling constant, and $\xi$ is randomly distributed in the sequence $[-1,1]$. Composite
disorder is realized by a random distribution of the propagation constants: $\kappa=\frac{\Delta\kappa}{2}\cdot\xi$, where $\Delta\kappa$ defines the
strength of the disorder in units of the propagation constant. Additionally, we assume periodic boundary conditions in the direction perpendicular to
the zigzag and bearded edges. In the upper row of Fig.~\ref{fig:eigval} we plot the eigenvalues in ascending order for no disorder
(Fig.~\ref{fig:eigval}(a), with a coupling constant $c_0=1\;cm^{-1}$), structural disorder (Fig.~\ref{fig:eigval}(c), $\Delta c = 8\; cm^{-1}$, $c_0=5\;cm^{-1}$, and fixed propagation constant $\kappa_0=0$) and composite disorder (Fig.~\ref{fig:eigval}(e), $\Delta\kappa = 8\;cm^{-1}$, $\kappa_0=0$, and fixed coupling constant $c_0=1\;cm^{-1}$). In the lower row of
Fig.~\ref{fig:eigval}, we plot the according DOS. In order to take into account the statistical nature of disorder \cite{Segev:Nature2007}, in both disorder
cases we averaged over $5000$ realizations, each containing $~750$ waveguides. As expected for graphene, the eigenvalue spectrum of the completely
ordered system shows a DOS equal to zero at $\beta=0$ (the Dirac points). In case of composite disorder the picture completely changes - the DOS is
not longer zero at $\beta=0$. Indeed, with structural disorder one still observes a vanishing DOS at $\beta=0$. Therefore, the chiral symmetry has
caused the preservation of the Dirac-like DOS even for very strong disorder.

\begin{figure}[htb]
\includegraphics[width=6.7cm]{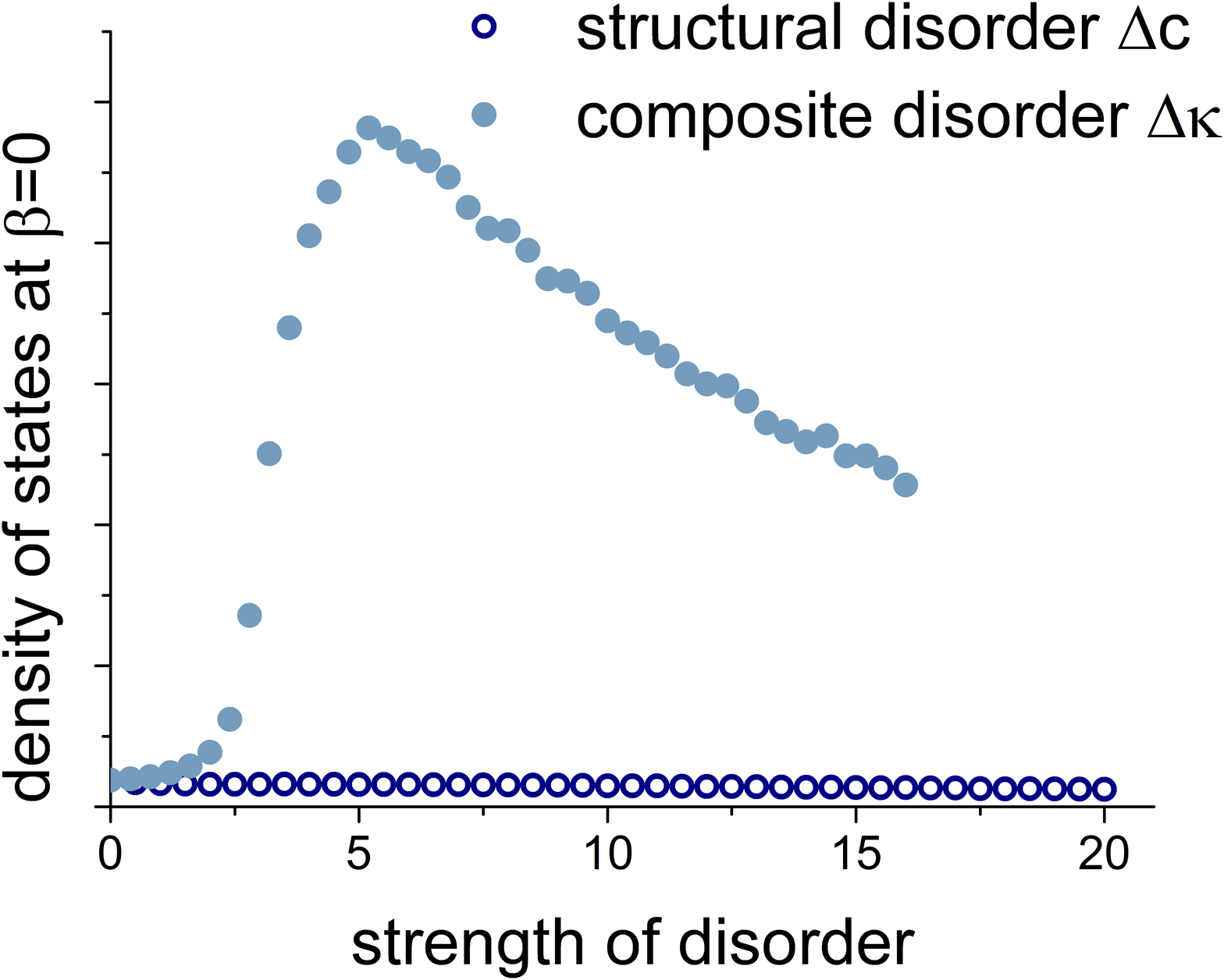}
\caption{Dependence of the DOS at $\beta=0$ on the strength of disorder plotted for structural and composite disorder. \label{fig:disstrength}}
\end{figure}

To further elaborate our results, in Fig.~\ref{fig:disstrength} we plot the simulated averaged DOS at $\beta=0$ for increasing composite and
structural disorder, respectively. The small fluctuations are due to the finite number of realizations, which were taken for the averaging process.
With increasing composite disorder the DOS at $\beta=0$ is increasing, so that the zero DOS disappears. In contrast, for increasing structural
disorder the DOS at $\beta=0$ stays constantly at zero since the chirality is preserved, even for very high disorder strengths. We attribute this
behavior to the intrinsically different impact of composite and structural disorder on the distribution of eigenvalues \cite{Lahini:PRA2011,Keil:OptLett2012}. For
structural disorder the eigenvalues appear in pairs with identical absolute values and opposite sign, such that they are symmetrically distributed
around $\beta=0$, which is not the case for composite disorder. This is a direct result of chiral symmetry breaking.

\begin{figure}[htb]
\includegraphics[width=6cm]{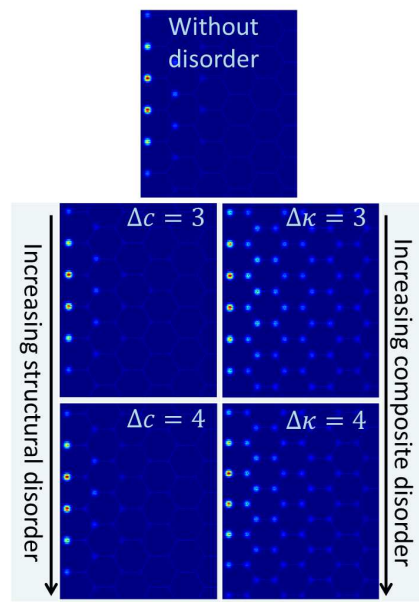}
\caption{Simulation of the light distribution after $z=7L_c$ when a broad Gaussian beam is launched at the armchair edge. The averaging was taken over
200 realizations. The edge state in the case of structural disorder propagates in a stable fashion, whereas it is destroyed for composite disorder.
\label{fig:simulation}}
\end{figure}

To test the conjectures based on the numerical simulations, we examine the \textit{edge states} of the honeycomb lattice. It is well known that in a
finite honeycomb lattice, edge states exist at the bearded and the zig-zag edges (for the nomenclature of the edges see Fig.~\ref{fig:setup}(a)).
Such states exist directly at $\beta=0$, and therefore in the same region as the Dirac points, where the DOS vanishes.  Probing the effects of
disorder on the edge states of the system represents an ideal diagnostic of the effects of chiral symmetry breaking and preservation (e.g., the effects on the DOS and the transport), for the following reasons.  In the case of composite disorder, zero-energy edge states may couple to bulk states
as a result of the disorder, and therefore spread into the bulk.  However, in the case of structural disorder, the chiral symmetry is preserved and
the edge states are not perturbed, and thus remain closely localized on the edge.  Thus, edge state properties yield a full description of Dirac
dynamics near $\beta=0$ under the influence of disorder. In this vein, the existence of an edge state can be used as an indicator whether the DOS vanishes at $\beta=0$ or not.

The predicted properties of the edge states under the influence of disorder are confirmed by tight-binding simulations of Eq.~\eqref{TBM}, which are
shown in Fig.~\ref{fig:simulation}. The bearded edge of the lattice is excited using an elliptical gaussian beam. Without disorder the edge state does not spread into the lattice. This is because it exists at the same energy as the Dirac
points where the DOS is zero, so that there are no states the edge state could couple to. If structural disorder (in the coupling coefficients) is
implemented the edge state is still present and can be observed even after a propagation length of $z=7L_c$ with the coupling length
$L_c=\frac{\pi}{2\bar{c}}$ and $\bar{c}$ as the mean coupling constant. In contrast, for composite disorder the edge state is destroyed and the light
spreads into the array as it is no longer localized at the edge. Note that in all cases, averaging was performed over 200 random realizations.

\begin{figure}[htb]
\includegraphics[width=6cm]{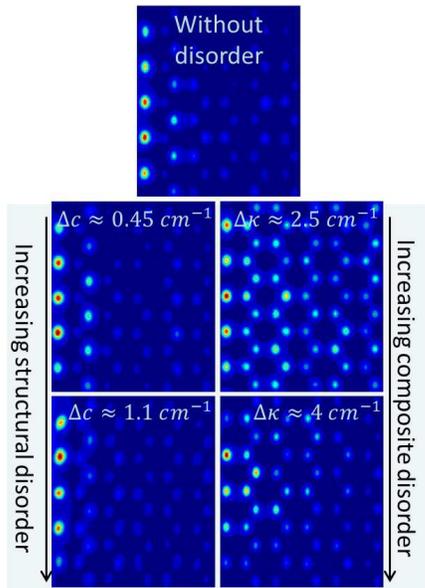}
\caption{Experimental light distribution at the end facet of the sample, when a broad edge state at the bearded edge is excited. The shown images are
averaged over 20 realizations. \label{fig:experiment}}
\end{figure}

In order to experimentally demonstrate the described effects, we fabricate a photonic graphene waveguide array using the femtosecond direct-writing
technology \cite{Szameit:PhDtutorial}. The structural disorder is realized by randomly shifting the positions of the waveguides in the transverse
directions with respect to their average positions by $\pm 1\;\mu m$ for a disorder of $\Delta c = 0.45\;cm^{-1}$, and $\pm 2\;\mu m$ for a disorder
of $\Delta c = 1.1\;cm^{-1}$, while the mean nearest-neighbor distance was $20\;\mu m$. The composite disorder was realized by changing the
translation velocity of the writing beam during fabrication. The mean velocity for all realizations was $80\;\frac{mm}{min}$ with a random shift of
$\pm 10\;\frac{mm}{min}$ and $\pm 15\;\frac{mm}{min}$ for $\Delta \kappa = 2.5\;cm^{-1}$ and $\Delta \kappa = 4.5\;cm^{-1}$, respectively. For the
investigation of the light propagation a broad elliptical beam with its main axis along the edge was launched into the bearded edge waveguides and
the output facet of the sample was imaged onto a CCD camera. For each disorder level, we averaged over 20 realizations, and the resulting averaged
output distributions are shown in Fig.~\ref{fig:experiment}. It is clearly seen that for increasing composite disorder the edge state couples to the
bulk modes and spreads into the lattice. Only for sufficiently high disorder Anderson localization takes place. In contrast, for structural disorder
our experiments unequivocally show that the edge state is stable and does not couple to the bulk modes. These results represent strong experimental
evidence that the chiral symmetry preserves the vanishing of the DOS at the Dirac point ($\beta=0$), whereas the breaking of chiral symmetry destroys
it.  Note that we checked experimentally that at disorder level of $\Delta c=0.45cm^{-1}$ no Anderson localization takes place.

In conclusion, we theoretically and experimentally analyzed the impact of composite and structural disorder on the DOS of photonic graphene. Our results are general and hold also for electronic graphene, where such controlled investigations of disorder are impossible.  We found that for
composite disorder, which breaks the chiral symmetry of the lattice, the DOS does not vanish at $\beta=0$. However, for structural disorder we could
experimentally demonstrate that edge states still exist, which implies that the DOS vanishes at $\beta=0$ even for strong disorder, so that chirality
prevails. Therefore, all properties of the graphene structure that arise from chirality and the vanishing DOS at $\beta=0$ do not break down for
structural disorder. Our work may provide further insight into certain properties of graphene that are strongly affected by disorder such as, e.g.,
the value of the minimal conductivity. Furthermore, our photonic system offers a unique platform for investigating the influence of
predetermined disorder on the electronic transport properties of graphene.

The authors thank the German Ministry of Education and Research for financial support (ZIK 03Z1HN31). M.C.R. is grateful to the Azrieli foundation
for the award of an Azrieli fellowship while at the Technion.


\begin{thebibliography}{26}
\expandafter\ifx\csname natexlab\endcsname\relax\def\natexlab#1{#1}\fi
\expandafter\ifx\csname bibnamefont\endcsname\relax
  \def\bibnamefont#1{#1}\fi
\expandafter\ifx\csname bibfnamefont\endcsname\relax
  \def\bibfnamefont#1{#1}\fi
\expandafter\ifx\csname citenamefont\endcsname\relax
  \def\citenamefont#1{#1}\fi
\expandafter\ifx\csname url\endcsname\relax
  \def\url#1{\texttt{#1}}\fi
\expandafter\ifx\csname urlprefix\endcsname\relax\def\urlprefix{URL }\fi
\providecommand{\bibinfo}[2]{#2}
\providecommand{\eprint}[2][]{\url{#2}}

\bibitem[{\citenamefont{Novoselov et~al.}(2004)\citenamefont{Novoselov, Geim,
  Morozov, Jiang, Zhang, Dubonos, Grigorieva, and
  Firsov}}]{Novolesov:Science2004}
\bibinfo{author}{\bibfnamefont{K.~S.} \bibnamefont{Novoselov}},
  \bibinfo{author}{\bibfnamefont{A.~K.} \bibnamefont{Geim}},
  \bibinfo{author}{\bibfnamefont{S.~V.} \bibnamefont{Morozov}},
  \bibinfo{author}{\bibfnamefont{D.}~\bibnamefont{Jiang}},
  \bibinfo{author}{\bibfnamefont{Y.}~\bibnamefont{Zhang}},
  \bibinfo{author}{\bibfnamefont{S.~V.} \bibnamefont{Dubonos}},
  \bibinfo{author}{\bibfnamefont{I.~V.} \bibnamefont{Grigorieva}},
  \bibnamefont{and} \bibinfo{author}{\bibfnamefont{A.~A.}
  \bibnamefont{Firsov}}, \bibinfo{journal}{Science}
  \textbf{\bibinfo{volume}{306}}, \bibinfo{pages}{666} (\bibinfo{year}{2004}).

\bibitem[{\citenamefont{Zhang et~al.}(2005)\citenamefont{Zhang, Tan, Stormer,
  and Kim}}]{Zhang:Nature2005}
\bibinfo{author}{\bibfnamefont{Y.}~\bibnamefont{Zhang}},
  \bibinfo{author}{\bibfnamefont{Y.-W.} \bibnamefont{Tan}},
  \bibinfo{author}{\bibfnamefont{H.~L.} \bibnamefont{Stormer}},
  \bibnamefont{and} \bibinfo{author}{\bibfnamefont{P.}~\bibnamefont{Kim}},
  \bibinfo{journal}{Nature} \textbf{\bibinfo{volume}{438}},
  \bibinfo{pages}{201} (\bibinfo{year}{2005}).

\bibitem[{\citenamefont{Novoselov et~al.}(2005)\citenamefont{Novoselov, Geim,
  Morozov, Jiang, Katsnelson, Grigorieva, Dubonos, and
  Firsov}}]{Novoselov:Nature2005}
\bibinfo{author}{\bibfnamefont{K.~S.} \bibnamefont{Novoselov}},
  \bibinfo{author}{\bibfnamefont{A.~K.} \bibnamefont{Geim}},
  \bibinfo{author}{\bibfnamefont{S.~V.} \bibnamefont{Morozov}},
  \bibinfo{author}{\bibfnamefont{D.}~\bibnamefont{Jiang}},
  \bibinfo{author}{\bibfnamefont{M.~I.} \bibnamefont{Katsnelson}},
  \bibinfo{author}{\bibfnamefont{I.~V.} \bibnamefont{Grigorieva}},
  \bibinfo{author}{\bibfnamefont{S.~V.} \bibnamefont{Dubonos}},
  \bibnamefont{and} \bibinfo{author}{\bibfnamefont{A.~A.}
  \bibnamefont{Firsov}}, \bibinfo{journal}{Nature}
  \textbf{\bibinfo{volume}{438}}, \bibinfo{pages}{197} (\bibinfo{year}{2005}).

\bibitem[{\citenamefont{Morozov et~al.}(2006)\citenamefont{Morozov, Novoselov,
  Katsnelson, Schedin, Ponomarenko, Jiang, and Geim}}]{Morozov:PRL2006}
\bibinfo{author}{\bibfnamefont{S.~V.} \bibnamefont{Morozov}},
  \bibinfo{author}{\bibfnamefont{K.~S.} \bibnamefont{Novoselov}},
  \bibinfo{author}{\bibfnamefont{M.~I.} \bibnamefont{Katsnelson}},
  \bibinfo{author}{\bibfnamefont{F.}~\bibnamefont{Schedin}},
  \bibinfo{author}{\bibfnamefont{L.~A.} \bibnamefont{Ponomarenko}},
  \bibinfo{author}{\bibfnamefont{D.}~\bibnamefont{Jiang}}, \bibnamefont{and}
  \bibinfo{author}{\bibfnamefont{A.~K.} \bibnamefont{Geim}},
  \bibinfo{journal}{Phys. Rev. Lett.} \textbf{\bibinfo{volume}{97}},
  \bibinfo{pages}{016801} (\bibinfo{year}{2006}).

\bibitem[{\citenamefont{Katsnelson et~al.}(2006)\citenamefont{Katsnelson,
  Novoselov, and Geim}}]{Katsnelson:NaturePhysics2006}
\bibinfo{author}{\bibfnamefont{M.~I.} \bibnamefont{Katsnelson}},
  \bibinfo{author}{\bibfnamefont{K.~S.} \bibnamefont{Novoselov}},
  \bibnamefont{and} \bibinfo{author}{\bibfnamefont{A.~K.} \bibnamefont{Geim}},
  \bibinfo{journal}{Nature Physics} \textbf{\bibinfo{volume}{2}},
  \bibinfo{pages}{620 } (\bibinfo{year}{2006}).

\bibitem[{\citenamefont{Wallace}(1947)}]{Wallace}
\bibinfo{author}{\bibfnamefont{P.~R.} \bibnamefont{Wallace}},
  \bibinfo{journal}{Phys. Rev.} \textbf{\bibinfo{volume}{71}},
  \bibinfo{pages}{622} (\bibinfo{year}{1947}).

\bibitem[{\citenamefont{Haldane}(1988)}]{Haldane}
\bibinfo{author}{\bibfnamefont{F.~D.~M.} \bibnamefont{Haldane}},
  \bibinfo{journal}{Phys. Rev. Lett.} \textbf{\bibinfo{volume}{61}},
  \bibinfo{pages}{2015} (\bibinfo{year}{1988}).

\bibitem[{\citenamefont{Hatsugai et~al.}(2006)\citenamefont{Hatsugai, Fukui,
  and Aoki}}]{Hatsugai2006}
\bibinfo{author}{\bibfnamefont{Y.}~\bibnamefont{Hatsugai}},
  \bibinfo{author}{\bibfnamefont{T.}~\bibnamefont{Fukui}}, \bibnamefont{and}
  \bibinfo{author}{\bibfnamefont{H.}~\bibnamefont{Aoki}},
  \bibinfo{journal}{Phys. Rev. B} \textbf{\bibinfo{volume}{74}},
  \bibinfo{pages}{205414} (\bibinfo{year}{2006}).

\bibitem[{\citenamefont{Hatsugai et~al.}(2007)\citenamefont{Hatsugai, Fukui,
  and Aoki}}]{Hatsugai2007}
\bibinfo{author}{\bibfnamefont{Y.}~\bibnamefont{Hatsugai}},
  \bibinfo{author}{\bibfnamefont{T.}~\bibnamefont{Fukui}}, \bibnamefont{and}
  \bibinfo{author}{\bibfnamefont{H.}~\bibnamefont{Aoki}}, \bibinfo{journal}{The
  European Physical Journal-Special Topics} \textbf{\bibinfo{volume}{148}},
  \bibinfo{pages}{133} (\bibinfo{year}{2007}).

\bibitem[{\citenamefont{Castro~Neto et~al.}(2009)\citenamefont{Castro~Neto,
  Guinea, Peres, Novoselov, and Geim}}]{GrapheneReview}
\bibinfo{author}{\bibfnamefont{A.~H.} \bibnamefont{Castro~Neto}},
  \bibinfo{author}{\bibfnamefont{F.}~\bibnamefont{Guinea}},
  \bibinfo{author}{\bibfnamefont{N.~M.~R.} \bibnamefont{Peres}},
  \bibinfo{author}{\bibfnamefont{K.~S.} \bibnamefont{Novoselov}},
  \bibnamefont{and} \bibinfo{author}{\bibfnamefont{A.~K.} \bibnamefont{Geim}},
  \bibinfo{journal}{Rev. Mod. Phys.} \textbf{\bibinfo{volume}{81}},
  \bibinfo{pages}{109} (\bibinfo{year}{2009}).

\bibitem[{\citenamefont{Altland}(2006)}]{Altland}
\bibinfo{author}{\bibfnamefont{A.}~\bibnamefont{Altland}},
  \bibinfo{journal}{Phys. Rev. Lett.} \textbf{\bibinfo{volume}{97}},
  \bibinfo{pages}{236802} (\bibinfo{year}{2006}).

\bibitem[{\citenamefont{Ostrovsky et~al.}(2006)\citenamefont{Ostrovsky, Gornyi,
  and Mirlin}}]{Merlin}
\bibinfo{author}{\bibfnamefont{P.~M.} \bibnamefont{Ostrovsky}},
  \bibinfo{author}{\bibfnamefont{I.~V.} \bibnamefont{Gornyi}},
  \bibnamefont{and} \bibinfo{author}{\bibfnamefont{A.~D.}
  \bibnamefont{Mirlin}}, \bibinfo{journal}{Phys. Rev. B}
  \textbf{\bibinfo{volume}{74}}, \bibinfo{pages}{235443}
  (\bibinfo{year}{2006}).

\bibitem[{\citenamefont{Das~Sarma et~al.}(2011)\citenamefont{Das~Sarma, Adam,
  Hwang, and Rossi}}]{DasSarma:RevModPhys2011}
\bibinfo{author}{\bibfnamefont{S.}~\bibnamefont{Das~Sarma}},
  \bibinfo{author}{\bibfnamefont{S.}~\bibnamefont{Adam}},
  \bibinfo{author}{\bibfnamefont{E.~H.} \bibnamefont{Hwang}}, \bibnamefont{and}
  \bibinfo{author}{\bibfnamefont{E.}~\bibnamefont{Rossi}},
  \bibinfo{journal}{Rev. Mod. Phys.} \textbf{\bibinfo{volume}{83}},
  \bibinfo{pages}{407} (\bibinfo{year}{2011}).

\bibitem[{\citenamefont{Mucciolo and Lewenkopf}(2010)}]{Mucciolo:JPhys2010}
\bibinfo{author}{\bibfnamefont{E.~R.} \bibnamefont{Mucciolo}} \bibnamefont{and}
  \bibinfo{author}{\bibfnamefont{C.~H.} \bibnamefont{Lewenkopf}},
  \bibinfo{journal}{Journal of Physics: Condensed Matter}
  \textbf{\bibinfo{volume}{22}}, \bibinfo{pages}{273201}
  (\bibinfo{year}{2010}).

\bibitem[{\citenamefont{Leconte et~al.}(2011)\citenamefont{Leconte, Lherbier,
  Varchon, Ordejon, Roche, and Charlier}}]{Leconte:PRB2011}
\bibinfo{author}{\bibfnamefont{N.}~\bibnamefont{Leconte}},
  \bibinfo{author}{\bibfnamefont{A.}~\bibnamefont{Lherbier}},
  \bibinfo{author}{\bibfnamefont{F.}~\bibnamefont{Varchon}},
  \bibinfo{author}{\bibfnamefont{P.}~\bibnamefont{Ordejon}},
  \bibinfo{author}{\bibfnamefont{S.}~\bibnamefont{Roche}}, \bibnamefont{and}
  \bibinfo{author}{\bibfnamefont{J.-C.} \bibnamefont{Charlier}},
  \bibinfo{journal}{Phys. Rev. B} \textbf{\bibinfo{volume}{84}},
  \bibinfo{pages}{235420} (\bibinfo{year}{2011}).

\bibitem[{\citenamefont{Wimmer et~al.}(2010)\citenamefont{Wimmer, Akhmerov, and
  Guinea}}]{Wimmer:PhysRevB2010}
\bibinfo{author}{\bibfnamefont{M.}~\bibnamefont{Wimmer}},
  \bibinfo{author}{\bibfnamefont{A.~R.} \bibnamefont{Akhmerov}},
  \bibnamefont{and} \bibinfo{author}{\bibfnamefont{F.}~\bibnamefont{Guinea}},
  \bibinfo{journal}{Phys. Rev. B} \textbf{\bibinfo{volume}{82}},
  \bibinfo{pages}{045409} (\bibinfo{year}{2010}).

\bibitem[{\citenamefont{Peleg et~al.}(2007)\citenamefont{Peleg, Bartal,
  Freedman, Manela, Segev, and Christodoulides}}]{Peleg:PRL2007}
\bibinfo{author}{\bibfnamefont{O.}~\bibnamefont{Peleg}},
  \bibinfo{author}{\bibfnamefont{G.}~\bibnamefont{Bartal}},
  \bibinfo{author}{\bibfnamefont{B.}~\bibnamefont{Freedman}},
  \bibinfo{author}{\bibfnamefont{O.}~\bibnamefont{Manela}},
  \bibinfo{author}{\bibfnamefont{M.}~\bibnamefont{Segev}}, \bibnamefont{and}
  \bibinfo{author}{\bibfnamefont{D.~N.} \bibnamefont{Christodoulides}},
  \bibinfo{journal}{Phys. Rev. Lett.} \textbf{\bibinfo{volume}{98}},
  \bibinfo{pages}{103901} (\bibinfo{year}{2007}).

\bibitem[{\citenamefont{Rechtsman
  et~al.}(2012{\natexlab{a}})\citenamefont{Rechtsman, Zeuner, T\"{u}nnermann,
  Nolte, Segev, and Szameit}}]{Rechtsman:NatPhoton2012}
\bibinfo{author}{\bibfnamefont{M.~C.} \bibnamefont{Rechtsman}},
  \bibinfo{author}{\bibfnamefont{J.~M.} \bibnamefont{Zeuner}},
  \bibinfo{author}{\bibfnamefont{A.}~\bibnamefont{T\"{u}nnermann}},
  \bibinfo{author}{\bibfnamefont{S.}~\bibnamefont{Nolte}},
  \bibinfo{author}{\bibfnamefont{M.}~\bibnamefont{Segev}}, \bibnamefont{and}
  \bibinfo{author}{\bibfnamefont{A.}~\bibnamefont{Szameit}},
  \bibinfo{journal}{Nat. Photon., advance online publication,
  doi:10.1038/nphoton.2012.302}  (\bibinfo{year}{2012}{\natexlab{a}}).

\bibitem[{\citenamefont{Kohmoto and Hasegawa}(2007)}]{Kohmoto:PhysRevB2007}
\bibinfo{author}{\bibfnamefont{M.}~\bibnamefont{Kohmoto}} \bibnamefont{and}
  \bibinfo{author}{\bibfnamefont{Y.}~\bibnamefont{Hasegawa}},
  \bibinfo{journal}{Phys. Rev. B} \textbf{\bibinfo{volume}{76}},
  \bibinfo{pages}{205402} (\bibinfo{year}{2007}).

\bibitem[{\citenamefont{Rechtsman
  et~al.}(2012{\natexlab{b}})\citenamefont{Rechtsman, Plotnik, Song, Heinrich,
  Zeuner, Nolte, Malkova, Xu, Szameit, Chen, and Segev}}]{Rechtsman:arxiv2012}
\bibinfo{author}{\bibfnamefont{M.~C.} \bibnamefont{Rechtsman}},
  \bibinfo{author}{\bibfnamefont{Y.}~\bibnamefont{Plotnik}},
  \bibinfo{author}{\bibfnamefont{D.}~\bibnamefont{Song}},
  \bibinfo{author}{\bibfnamefont{M.}~\bibnamefont{Heinrich}},
  \bibinfo{author}{\bibfnamefont{J.~M.} \bibnamefont{Zeuner}},
  \bibinfo{author}{\bibfnamefont{S.}~\bibnamefont{Nolte}},
  \bibinfo{author}{\bibfnamefont{N.}~\bibnamefont{Malkova}},
  \bibinfo{author}{\bibfnamefont{J.}~\bibnamefont{Xu}},
  \bibinfo{author}{\bibfnamefont{A.}~\bibnamefont{Szameit}},
  \bibinfo{author}{\bibfnamefont{Z.}~\bibnamefont{Chen}}, \bibnamefont{and}
  \bibinfo{author}{\bibfnamefont{M.}~\bibnamefont{Segev}},
  \bibinfo{journal}{arXiv:1210.5361 [cond-mat.mes-hall]}
  (\bibinfo{year}{2012}{\natexlab{b}}).

\bibitem[{\citenamefont{Liu et~al.}(2009)\citenamefont{Liu, Suenaga, Harris,
  and Iijima}}]{Liu:PRL2009}
\bibinfo{author}{\bibfnamefont{Z.}~\bibnamefont{Liu}},
  \bibinfo{author}{\bibfnamefont{K.}~\bibnamefont{Suenaga}},
  \bibinfo{author}{\bibfnamefont{P.~J.~F.} \bibnamefont{Harris}},
  \bibnamefont{and} \bibinfo{author}{\bibfnamefont{S.}~\bibnamefont{Iijima}},
  \bibinfo{journal}{Phys. Rev. Lett.} \textbf{\bibinfo{volume}{102}},
  \bibinfo{pages}{015501} (\bibinfo{year}{2009}).
  
 \bibitem[{\citenamefont{Rechtsman
  et~al.}(2012{\natexlab{b}})\citenamefont{Rechtsman, Plotnik,
  Zeuner, Szameit, and Segev}}]{strainedHC}
\bibinfo{author}{\bibfnamefont{M.~C.} \bibnamefont{Rechtsman}},
  \bibinfo{author}{\bibfnamefont{Y.}~\bibnamefont{Plotnik}},
  \bibinfo{author}{\bibfnamefont{J.~M.} \bibnamefont{Zeuner}},
  \bibinfo{author}{\bibfnamefont{A.}~\bibnamefont{Szameit}}, \bibnamefont{and}
  \bibinfo{author}{\bibfnamefont{M.}~\bibnamefont{Segev}},
  \bibinfo{journal}{arXiv:1211.5683 [cond-mat.mes-hall]}
  (\bibinfo{year}{2012}{\natexlab{b}}).
 
  \bibitem[{\citenamefont{Bellec et~al.}(2013)\citenamefont{Bellec, Kuhl, Montambaux, and Mortessagne}}]{Bellec:PRL2013}
\bibinfo{author}{\bibfnamefont{M.}~\bibnamefont{Bellec}},
  \bibinfo{author}{\bibfnamefont{U.}~\bibnamefont{Kuhl}},
  \bibinfo{author}{\bibfnamefont{G.} \bibnamefont{Montambaux}},
  \bibnamefont{and} \bibinfo{author}{\bibfnamefont{F.}~\bibnamefont{Mortessagne}},
  \bibinfo{journal}{Phys. Rev. Lett.} \textbf{\bibinfo{volume}{110}},
  \bibinfo{pages}{033902} (\bibinfo{year}{2013}).

\bibitem[{\citenamefont{Geim and Novoselov}(2007)}]{Geim:NatMat2007}
\bibinfo{author}{\bibfnamefont{A.~K.} \bibnamefont{Geim}} \bibnamefont{and}
  \bibinfo{author}{\bibfnamefont{K.~S.} \bibnamefont{Novoselov}},
  \bibinfo{journal}{Nat. Mater.} \textbf{\bibinfo{volume}{6}},
  \bibinfo{pages}{183} (\bibinfo{year}{2007}).

\bibitem[{\citenamefont{Schwartz et~al.}(2007)\citenamefont{Schwartz, Bartal,
  Fishman, and Segev}}]{Segev:Nature2007}
\bibinfo{author}{\bibfnamefont{T.}~\bibnamefont{Schwartz}},
  \bibinfo{author}{\bibfnamefont{G.}~\bibnamefont{Bartal}},
  \bibinfo{author}{\bibfnamefont{S.}~\bibnamefont{Fishman}}, \bibnamefont{and}
  \bibinfo{author}{\bibfnamefont{M.}~\bibnamefont{Segev}},
  \bibinfo{journal}{Nature} \textbf{\bibinfo{volume}{446}}, \bibinfo{pages}{52}
  (\bibinfo{year}{2007}).

\bibitem[{\citenamefont{Lahini et~al.}(2011)\citenamefont{Lahini, Bromberg,
  Shechtman, Szameit, Christodoulides, Morandotti, and
  Silberberg}}]{Lahini:PRA2011}
\bibinfo{author}{\bibfnamefont{Y.}~\bibnamefont{Lahini}},
  \bibinfo{author}{\bibfnamefont{Y.}~\bibnamefont{Bromberg}},
  \bibinfo{author}{\bibfnamefont{Y.}~\bibnamefont{Shechtman}},
  \bibinfo{author}{\bibfnamefont{A.}~\bibnamefont{Szameit}},
  \bibinfo{author}{\bibfnamefont{D.~N.} \bibnamefont{Christodoulides}},
  \bibinfo{author}{\bibfnamefont{R.}~\bibnamefont{Morandotti}},
  \bibnamefont{and}
  \bibinfo{author}{\bibfnamefont{Y.}~\bibnamefont{Silberberg}},
  \bibinfo{journal}{Phys. Rev. A} \textbf{\bibinfo{volume}{84}},
  \bibinfo{pages}{041806} (\bibinfo{year}{2011}).

\bibitem[{\citenamefont{Keil et~al.}(2012)\citenamefont{Keil, Lahini,
  Shechtman, Heinrich, Pugatch, Dreisow, T\"{u}nnermann, Nolte, and
  Szameit}}]{Keil:OptLett2012}
\bibinfo{author}{\bibfnamefont{R.}~\bibnamefont{Keil}},
  \bibinfo{author}{\bibfnamefont{Y.}~\bibnamefont{Lahini}},
  \bibinfo{author}{\bibfnamefont{Y.}~\bibnamefont{Shechtman}},
  \bibinfo{author}{\bibfnamefont{M.}~\bibnamefont{Heinrich}},
  \bibinfo{author}{\bibfnamefont{R.}~\bibnamefont{Pugatch}},
  \bibinfo{author}{\bibfnamefont{F.}~\bibnamefont{Dreisow}},
  \bibinfo{author}{\bibfnamefont{A.}~\bibnamefont{T\"{u}nnermann}},
  \bibinfo{author}{\bibfnamefont{S.}~\bibnamefont{Nolte}}, \bibnamefont{and}
  \bibinfo{author}{\bibfnamefont{A.}~\bibnamefont{Szameit}},
  \bibinfo{journal}{Opt. Lett.} \textbf{\bibinfo{volume}{37}},
  \bibinfo{pages}{809} (\bibinfo{year}{2012}).

\bibitem[{\citenamefont{Szameit and Nolte}(2010)}]{Szameit:PhDtutorial}
\bibinfo{author}{\bibfnamefont{A.}~\bibnamefont{Szameit}} \bibnamefont{and}
  \bibinfo{author}{\bibfnamefont{S.}~\bibnamefont{Nolte}},
  \bibinfo{journal}{Journal of Physics B: Atomic, Molecular and Optical
  Physics} \textbf{\bibinfo{volume}{43}}, \bibinfo{pages}{163001}
  (\bibinfo{year}{2010}).

\end{thebibliography}

\end{document}